\documentstyle[12pt,epsf,epsfig]{article}
\newcommand{\beq}{\begin{equation}}
\newcommand{\eeq}{\end{equation}}
\newcommand{\bea}{\begin{eqnarray}}
\newcommand{\eea}{\end{eqnarray}}

\def\lsim{\mathrel{\vcenter{\hbox{$<$}\nointerlineskip\hbox{$\sim$}}}}

\def\m12{m_{1\!/2}}

\begin{document}
\begin{titlepage}
\pagestyle{empty}
\baselineskip=21pt
\rightline{hep-ph/0206110}
\rightline{CERN--TH/2002-126}     
\rightline{ICRR-REPORT-490-2002-8}
\rightline{DPNU-02-16}
\vskip 0.25in
\begin{center}
{\large{\bf 
A New Parametrization of the Seesaw Mechanism and   \\
Applications in Supersymmetric Models
}}
\end{center}
\begin{center}
\vskip 0.25in
{{\bf John Ellis}$^1$,
{\bf Junji Hisano}$^{2}$,
{\bf Martti Raidal}$^{1,3}$ and 
{\bf Yasuhiro Shimizu}$^{4}$
\vskip 0.15in
{\it
$^1${CERN, CH 1211 Geneva 23, Switzerland}\\
$^2${ICRR, University of Tokyo, Kashiwa 277-8582, Japan }\\
$^3${National Institute of Chemical Physics and Biophysics, \\ 
Tallinn 10143, Estonia }\\
$^4${Department of Physics, Nagoya University, Nagoya 464-8692, Japan}\\
}}
\vskip 0.25in
{\bf Abstract}
\end{center}
\baselineskip=18pt \noindent

We present a new parametrization of the minimal seesaw model,
expressing the heavy-singlet neutrino Dirac Yukawa couplings
$(Y_\nu)_{ij}$ and Majorana masses $M_{N_i}$ in terms of effective
light-neutrino observables and an auxiliary Hermitian matrix $H.$ In
the minimal supersymmetric version of the seesaw model, the latter can
be related directly to other low-energy observables, including
processes that violate charged lepton flavour and CP. This
parametrization enables one to respect the stringent constraints on
muon-number violation while studying the possible ranges for other
observables by scanning over the allowed parameter space of the model.
Conversely, if any of the lepton-flavour-violating process is
observed, this measurement can be used directly to constrain
$(Y_\nu)_{ij}$ and $M_{N_i}.$ As applications, we study flavour-violating
$\tau$ decays and the electric dipole moments of leptons in the
minimal supersymmetric seesaw model.

\vfill
\leftline{CERN--TH/2002-126}
\leftline{June 2002}
\end{titlepage}
\baselineskip=18pt


\section{Introduction}

Experiments on both atmospheric and solar neutrinos have now provided
`smoking guns' for neutrino oscillations. Most recently, the direct SNO
measurement of the solar neutrino flux via neutral-current scattering
confirms solar neutrino oscillations~\cite{Ahmad:2002jz} and favours
strongly the LMA solution~\cite{Ahmad:2002ka}. This region of parameter
space is within reach of the KamLAND experiment, and is expected to be
probed soon \cite{kamland}. The existence of large mixing angles for both
solar and atmospheric neutrinos~\cite{skatm} is one of the biggest
mysteries in particle physics. The most favoured mechanism for generating
neutrino masses is the seesaw mechanism \cite{seesaw}, which naturally
explains their small sizes. However, it is an open question whether the
seesaw mechanism can explain why mixing in the lepton sector seems to be
larger than in the quark sector. In the absence of a theory of flavour, it
is important to study the consequences of neutrino mixing for as many
physical observables as possible.

In the minimal supersymmetric seesaw model, lepton-flavor-violating (LFV)
phenomena provide a tool to study indirectly neutrino parameters and probe
other aspects beyond the large mixing angles measured in neutrino
oscillations. If supersymmetry breaking originates from physics beyond the
heavy singlet neutrino mass scale, LFV slepton masses are induced
radiatively~\cite{bm,Hisano:1995nq} via the Dirac Yukawa couplings of the
neutrinos, even if the input supersymmetry-breaking parameters are
flavor-blind. On the other hand, the light neutrino masses and mixings
depend on both the Yukawa couplings and the Majorana masses of the heavy
singlet neutrinos. Thus one can hope to reconstruct the physical
parameters in the heavy singlet-neutrino sector entirely in terms of the
light neutrino data and low-energy observables such as rates for LFV
processes \cite{di}. 
To this end, in this paper we present a parameterization of the
minimal seesaw model and apply it to the minimal supersymmetric version of
the seesaw model.

The essence of our parametrization is the following.  The minimal seesaw
mechanism, whether supersymmetric or non-supersymmetric, involves 18
physical degrees of freedom, including 6 real mixing angles and 6
CP-violating phases. On the other hand, the induced light-neutrino mass
matrix has 9 degrees of freedom, including 3 real mixing angles and 3
CP-violating phases. Thus we need 9 additional degrees of freedom to
parametrize completely the seesaw mechanism. These can be chosen in such a
way as to be related to low-energy leptonic observables in the
supersymmetric version of the seesaw model. We recall that the LFV
renormalization of the supersymmetry-breaking parameters at low energy are
proportional to
\begin{eqnarray}
H_{ij}
&=&
\sum_k 
{(Y_\nu^*)_{ki}}
{(Y_\nu)_{kj}}
\log\frac{M_G}{M_{N_k}}, 
\end{eqnarray}
where $(Y_\nu)$ and ${M}_{N}$ are the heavy singlet-neutrino Dirac
Yukawa couplings and Majorana masses, respectively, and $M_G$ is the GUT
scale where the initial conditions for the supersymmetry-breaking
parameters are imposed. Since $H$ is a Hermitian matrix, it has 9 degrees
of freedom including 3 real mixings and 3 phases. This implies that we can
parametrize the seesaw mechanism by the light neutrino mass matrix ${\cal
M}_\nu$ and the Hermitian matrix $H$ according to
\bea
({\cal M}_\nu\,, H)\longrightarrow (Y_\nu \,,{M}_{N})\,.
\eea 
As a result, we can obtain $Y_\nu$ and ${M}_{N}$ that yield
automatically the light neutrino masses and mixings measured in
oscillation experiments. However, the main motivation for our
parametrization comes from its power in studies of the charged-lepton
physics in the supersymmetric seesaw model~\footnote{We emphasize, though,
that the parametrization itself is more general, and does not depend on
the existence of supersymmetry.}.

The LMA solution to the solar neutrino anomaly tends to predict a large
branching ratios for $\mu\rightarrow e \gamma$ in the supersymmetric
seesaw model~\cite{hn,ci,nlfv}, which may be within reach of near-future
experiments, or even beyond the current experimental bound~\footnote{Also,
some explicit models predict the third neutrino mixing parameter $U_{e3}$
to be ${\cal O}(10^{-(1-2)})$, which may also lead to a large branching ratio for
$\mu\rightarrow e \gamma$ \cite{tobe}.}.  This does not imply that the
supersymmetric seesaw model is strongly constrained, because it has a
multi-dimensional parameter space. However, it is difficult to scan
efficiently over the allowed parameter space while satisfying the
$\mu\rightarrow e \gamma$ constraint. Our parametrization solves this
difficulty, because the parameter matrix $H$ is related to the solutions
of the renormalization-group equations. Therefore, it is straightforward
to choose a parameter region where $\mu\rightarrow e \gamma$ is
suppressed, but the other low-energy observables may vary over their full
ranges. Furthermore, if some future experiment discovers a LFV process or
the electric dipole moment (EDM) of some lepton, this observation will be
directly related to $H$ and thus to the neutrino parameters. In our
parametrization, the high-energy neutrino couplings and masses can be
expressed entirely in terms of the induced low-energy observables.

Our work is organized as follows. In Section 2 we outline the new
parametrization and our procedure for analyzing charged-lepton decays. In
Section 3 we explain the relation between our parametrization and the
physical observables. In Section 4 we present a study of LFV $\tau$ decays
and the EDMs of the electron and muon in the supersymmetric seesaw model,
as applications of our approach. We find that $\tau\rightarrow \mu(e)  
\gamma$ can saturate the current experimental bound, even when
$\mu\rightarrow e \gamma$ is suppressed enough to be acceptable.  
The EDMs of the muon and
electron generally fall below $10^{-27}(10^{-29})e$~cm in our random
parameter scan. We also present the relation between
Br($\tau\rightarrow \mu(e)\gamma$) and Br($\tau\rightarrow
\mu(e)\ell^+\ell^-$). Section 5 summarizes our conclusions.

\section{Parametrization of Neutrino Couplings and Masses }

In view of the subsequent application to the supersymmetric version of the 
seesaw model, we illustrate the parametrization for this case, though it 
is also valid in the absence of supersymmetry. The leptonic superpotential 
of the supersymmetric version of the minimal seesaw model is
\begin{eqnarray}
W = N^{c}_i (Y_\nu)_{ij} L_j H_2
  + E^{c}_i (Y_e)_{ij}  L_j H_1 
  + \frac{1}{2}{N^c}_i (M_N)_{ij} N^c_j \,,
\end{eqnarray}
where the indices $i,j$ run over three generations and ${(M_N)}_{ij}$
is the heavy singlet-neutrino mass matrix.  In addition to the three 
charged-lepton masses, this superpotential has 18
physical parameters, including 6 real mixing angles and 6 CP-violating phases. 

At low energies, the effective superpotential obtained by integrating out 
the heavy neutrinos is
\begin{eqnarray}
W_{\rm eff} =  E^{c}_i (Y_e)_{i}  L_j H_1 
  + \frac{1}{2 v^2 \sin^2\beta} ({\cal M_\nu})_{ij} (L_i H_2)(L_j H_2) \,,
\label{weff}
\end{eqnarray}
where we work in a basis in which the charged-lepton Yukawa couplings are
diagonal. The second term in (\ref{weff}) leads to the light neutrino 
masses and mixings. The explicit form of  ${\cal M_\nu}$ is given by
\begin{eqnarray}
({\cal M_\nu})_{ij} &=&
\sum_k \frac{(Y_\nu({Q}_k))_{ki}(Y_\nu({Q}_k))_{kj}}
            {{M}_{N_k}} \,,
\label{lightMnu}
\end{eqnarray}
where the heavy-singlet neutrino Dirac Yukawa couplings $Y_\nu$ and masses 
${M}_{N_i}$ are defined at the renormalization scale 
$Q_k=M_{N_k}$, and in our notation ${M}_{N_1}<{M}_{N_2}<{M}_{N_3}$. It is important to distinguish between the renormalization scales
for different components in the Yukawa coupling matrix,
since the EDMs of charged leptons in the supersymmetric seesaw model
are sensitive to non-universal radiative corrections to the 
supersymmetry-breaking parameters, which come from the
non-degeneracy of the heavy singlet neutrino masses~\cite{Ellis:2001yz}. 
For simplicity, we ignore the
renormalization of  ${\cal M}_\nu$ after the decoupling of the singlet
neutrinos.

The light neutrino mass matrix ${\cal
M_\nu}$ (\ref{lightMnu}) is symmetric, with 9 parameters,
including 3 real
mixing angles and 3 CP-violating phases. It can be diagonalized by a 
unitary
matrix $U$ as
\begin{eqnarray}
U^T {\cal M}_\nu U &=& {\cal M}_\nu^D\,.
\end{eqnarray}
By redefinition of fields one can rewrite $U \equiv V P,$ where
$P \equiv {\rm diag}(e^{i\phi_1}, e^{i\phi_2}, 1 )$ and $V$ is the MNS 
matrix, with the 3 real mixing angles and the remaining CP-violating 
phase. \\

The key proposal of this paper is to characterize the seesaw 
neutrino sector by
${\cal M}_\nu$ and a Hermitian matrix $H$, whose diagonal terms are
real and positive, which is defined in terms of $Y_\nu$ and the heavy 
neutrino masses ${M}_{N}$ by
\begin{eqnarray}
H_{ij}
&=&
\sum_k 
{(Y_\nu^*({Q}_k))_{ki}}
{(Y_\nu({Q}_k))_{kj}}
\log\frac{M_G}{{M}_{N_k}}, 
\end{eqnarray}
with $M_G$ the GUT scale. The Hermitian matrix $H$ has 9 parameters 
including 3 phases, which are clearly independent of the parameters
in ${\cal M}_\nu$. Thus ${\cal M}_\nu$ and $H$ together provide the 
required 18 parameters, including 6 CP-violating phases. 

Although our parametrization also includes an unphysical region, it has
the merit of suitability for comprehensive studies of the minimal
supersymmetric seesaw model. In this model, the non-universal elements in
the left-handed slepton mass matrix, which induce the charged LFV
observables, are approximately proportional to $H$ if the slepton masses
are flavour independent at $M_G$. Thus, this parameterization allows us to
control the LFV processes and scan over the allowed parameter space at the
same time. Conversely, if some LFV process is discovered in the future,
its measurement can be incoprorated directly into our parametrization of
the neutrino sector.

We now explain how to reconstruct the heavy singlet-neutrino sector
from knowledge of ${\cal M}_\nu$ and $H$. First we recall the 
parametrization of the neutrino Dirac Yukawa coupling given in~\cite{ci},
\begin{eqnarray}
(Y_\nu ({Q}_i))_{ij}
&=& 
\left.
\frac{\sqrt{{M_N}} R \sqrt{{\cal M}_\nu}\, U^\dagger}
     {v\sin\beta}
\right|_{ij},
\label{yci} 
\end{eqnarray}
where $R$ is an auxiliary complex orthogonal matrix: $RR^T=R^TR=1$.  
Using this parametrization, $H$ becomes
\begin{eqnarray}
H&=& 
\frac{1}{v^2\sin^2\beta}
U\,\sqrt{{\cal M}_\nu} R^\dagger 
\overline{{M_N}} 
R \sqrt{{\cal M}_\nu}\, U^\dagger
\end{eqnarray}
where $\overline{{M_{N_i}}} \equiv {M_{N_i}}
\log({M_G}/{{{M}_{N_i}}})$. If we can diagonalize the following
Hermitian matrix $H',$
\begin{eqnarray}
H' &=& \sqrt{{\cal M}_\nu}^{-1} U^\dagger H U \sqrt{{\cal M}_\nu}^{-1} 
\, v^2 \sin^2\beta ,
\end{eqnarray}
by the complex orthogonal matrix $R'$:
\bea
H'={R'}^\dagger \overline{{M}_{N}}R',
\eea
then we can calculate the heavy singlet neutrino masses from 
$\overline{{M}_{N}}$ and the corresponding 
$Y_\nu$ from (\ref{yci}) taking $R=R'$.

However, the Hermitian matrix $H'$ cannot always be diagonalized by a
complex orthogonal matrix: the condition for such a diagonalization is
that all the eigenvalues of ${H'}^{\star}H'$ are positive, in which case
$R'$ is given by the eigenvectors of ${H'}^{\star}H'$.  This reflects the
fact that our parametrization also includes an unphysical region, so that
every chosen $H$ does not necessarily give physical neutrino masses and
couplings. Since our objective in this paper is to survey the
multi-dimensional parameter space using scatter plots, this shortcoming is
not critical.

In our subsequent analysis, we first generate randomly the matrix $H$, the
phases and the common mass scale in the light neutrino sector, and then
calculate the corresponding heavy neutrino masses and couplings. The
Yukawa couplings $(Y_\nu)_{ij}$ contribute to the renormalization-group
(RG) equations above ${M}_{N_i}$, since the corresponding singlet
neutrino is dynamical there. When we derive the Yukawa couplings at the
GUT scale, we introduce $(Y_\nu)_{ij}$ in the RG equations at $Q_i={
M}_{N_i}$ where the neutrinos appear. When evaluating the
supersymmetry-breaking parameters at the weak scale, the right-handed
neutrinos are integrated out at their own mass scales.

\section{Observables}

In the previous Section we presented our parametrization of the 
minimal seesaw mechanism in terms of the light-neutrino mass matrix
${\cal M}_{\nu}$ and a Hermitian parameter matrix $H.$ 
Here we make explicit the correspondence between
this parametrization and low-energy observables in the 
supersymmetric version of the seesaw model. 

\subsection{Neutrino Experiments}
As already mentioned, the light-neutrino mass matrix ${\cal M}_\nu$ 
contains nine physical
parameters: 3 mass eigenvalues, 3 mixing angles, 1 CP-violating mixing
phase in the MNS matrix, and 2 CP-violating Majorana phases,
the LMA solution is now favoured, following the SNO 
neutral-current result. Thus, the favoured regions for
the atmospheric and solar neutrino parameters are
\begin{eqnarray}
\Delta m^2_{32} &=& (1-5)\times 10^{-3} {\rm eV^2} \,,\\
\sin^2 2 \theta_{23} &=& (0.8-1.0) \,,\\
\Delta m^2_{21} &=& 10^{-(4-5)}{\rm eV^2} \,,\\
\tan^2 \theta_{12} &\simeq& (0.2-0.6) \,.
\end{eqnarray}
The CHOOZ~\cite{CHOOZ} and Palo Verde~\cite{Boehm:2001ik} experiments
provide the constraint
\begin{eqnarray}
\sin^2 2 \theta_{13} &\lsim& 0.1 \,.
\end{eqnarray}
These parameters, together with the CP-violating mixing phase in the MNS
matrix, may be measured in future experiments, such as
the KamLAND and the neutrino factory. There would still be three
undetermined parameters, the normalization of the neutrino mass and
the Majorana phases. The neutrinoless double beta decay matrix element is
proportional to
\begin{eqnarray}
|m_{ee}|=\left|\sum_i U_{ei}^* m_{\nu_i} U_{ie}^\dagger \right|\,,
\end{eqnarray}
and so would provide a constraint on the neutrino mass scale and Majorana 
phases, if it could be measured.

\subsection{Charged LFV Processes}
If the supersymmetry-breaking parameters at the GUT scale are universal, 
off-diagonal components in the left-handed slepton mass matrix $m_{\tilde
L}$ and the trilinear slepton coupling $A_e$ are 
induced by renormalization, taking the approximate forms
\begin{eqnarray}
(\delta m_{\tilde{L}}^2)_{ij}&\simeq&
-\frac{1}{8\pi^2}(3m_0^2+A_0^2) H_{ij} \,,
\nonumber\\
(\delta A_e)_{ij} &\simeq&
-\frac{1}{8\pi^2} A_0 Y_{e_i} H_{ij} \,,
\label{leading}
\end{eqnarray}
where $ i\ne j$, and the off-diagonal components of the right-handed
slepton mass matrix are suppressed. The parameters $m_0$ and $A_0$ are
the universal scalar mass and trilinear coupling at the GUT scale.
Here, we ignore terms of higher order in $Y_e$, assuming that $\tan\beta$
is not extremely large.  Thus, the parameters in $H$ may in principle be 
determined by the LFV processes of charged leptons. Currently,
$\mu\rightarrow e \gamma$ experiments give the following
constraints on them:
\begin{eqnarray}
H_{12}&\lsim& 10^{-2} \times \tan^{-\frac12} \beta 
              \left(\frac{m_0}{100{\rm GeV}}\right)^2
              \left(
              \frac{{\rm Br}(\mu\rightarrow e\gamma)}
                   {1.2\times 10^{-11}}
              \right)^{\frac{1}{2}}\,,
\nonumber\\
H_{13}H_{32}&\lsim& 10^{-1} \times \tan^{-\frac12} \beta 
              \left(\frac{m_0}{100{\rm GeV}}\right)^2
              \left(
              \frac{{\rm Br}(\mu\rightarrow e\gamma)}
                   {1.2\times 10^{-11}}
              \right)^{\frac{1}{2}}\,,
\label{mueg_const}
\end{eqnarray}
where we take $(m^2_{\tilde{L}})_{ii}\sim m_0^2$. These components may 
also be measured directly in future collider experiments, if the sleptons 
are produced there~\cite{lfvcp,collider}.

Although the matrix $H$ has three CP-violating phases, two of them are
almost irrelevant to charged LFV phenomena. The two phases may be
moved from $H$ to ${\cal M}_\nu$ by a rotation of $L$. In fact,
there is only a single Jarlskog invariant obtainable from $H$~\cite{lfvcp}:
\begin{eqnarray}
J&=&{\rm Im} H_{12} H_{23} H_{31}\,,
\end{eqnarray}
which determines the T-odd asymmetry in $\mu\rightarrow 3e$~\cite{3l}.

We kept in (\ref{leading}) only the leading-order contributions
to the soft supersymmetry-breaking parameters, and ignored higher-order
corrections. If some components of $H$ are suppressed, non-trivial
flavour structure may emerge in the higher-order corrections. At
${\cal O}(\log^2M_G/{M}_{N_3})$ or ${\cal O}(\log M_G/{M}_{N_3}\log{
M}_{N_j}/{M}_{N_i})$ $(i\ne j)$, $(m_{\tilde{L}}^2)$ and $(A_e)$
get the following corrections:
\begin{eqnarray}
(\delta' m_{\tilde{L}}^2)_{ij}&\simeq&
\frac{1}{(4\pi)^4}
(
A_0^2 H^2)_{ij}
\nonumber\\
&&
+\frac{6}{(4\pi)^4}
 (3 m_0-A_0^2)
\sum_{k < l}
\left\{X_k,X_l\right\}\log\frac{{M}_{N_l}}{{M}_{N_k}}
\log\frac{M_G}{{M}_{N_3}}
)_{ij}
 \,,
\label{thesm}\\
(\delta' A_e)_{ij} &\simeq&
\frac{1}{(4\pi)^4}  A_0 Y_{e_i}
\nonumber\\
&&\times
\sum_{k < l}
\left[
6\left\{X_k,X_l\right\} \log\frac{{M}_{N_l}}{{M}_{N_k}}
+4\left[X_k,X_l\right] \log\frac{{M}_{N_l}}{{M}_{N_k}}
\right]_{ij}\log\frac{M_{G}}{{M}_{N_3}} \,,
\label{athre}
\end{eqnarray}
where
\begin{eqnarray}
(X_k)_{ij}  &=&  
{(Y_\nu^*({M}_{N_k}))_{ki}}
{(Y_\nu({M}_{N_k}))_{kj}}\,.
\end{eqnarray}
The second term in (\ref{thesm}) and the term in (\ref{athre})
come from threshold corrections at the heavy singlet-neutrino scale,
due to the non-degeneracy of their neutrino masses. 

\subsection{EDMs of the Charged Leptons}

The EDMs of the charged leptons depend non-trivially on the parameters in 
the supersymmetric seesaw model. If the heavy singlet-neutrino masses are
degenerate, the EDMs of the charged leptons are strongly suppressed
by the chiral structure of the seesaw model, and are proportional to
\begin{eqnarray}
{\rm Im} \left[\left[Y_e Y_\nu^\dagger Y_\nu 
\left[Y_e^\dagger Y_e,~ Y_\nu^\dagger Y_\nu \right] 
Y_\nu^\dagger Y_\nu \right]_{ii}\right] \,. 
\end{eqnarray}
This is similar to the neutron EDM in the Standard Model. However, when 
the heavy singlet-neutrino
masses are not degenerate, the EDMs may be enhanced significantly. The
trilinear coupling $A_e$ gets a threshold correction at the
heavy singlet neutrino scale, and may get radiatively-induced
diagonal phases  proportional to 
\begin{eqnarray}
{\rm Im}[X_j,X_k]_{ii} \log{{M}_{N_k}}/{{M}_{N_j}}&\ne&0,
\nonumber
\end{eqnarray}
as in (\ref{athre})~\cite{Ellis:2001yz}. This depends non-trivially
on the CP-violating phases, including the two Majorana phases in
${\cal M}_\nu$ and two phases in $H$ that are irrelevant for LFV.

\section{Phenomenological Analysis}

Using the parametrization proposed above, we now study the branching
ratios for LFV $\tau$ decays, such as Br$(\tau\to \mu\gamma)$, Br$(\tau\to
e\gamma)$ and Br$(\tau\to 3l)$, and the EDMs $d_\mu$ and $d_e$. We fix the
light neutrino parameters:  $\Delta m^2_{32}=3\times 10^{-3}$ eV$^2,$
$\Delta m^2_{21}=4.5\times 10^{-5}$ eV$^2,$ $\tan^2\theta_{23}=1$ and
$\tan^2\theta_{12}=0.4$ corresponding to the LMA solution for the solar
neutrino anomaly. Since the bound on the angle $\theta_{13}$ is quite
stringent, our results depend very weakly on its actual value. We fix
$\sin\theta_{13}=0.1$ and $\delta=\pi/2$. We study both the normal and the
inverse hierarchy of light neutrino masses, since neutrino oscillations do
not discriminate between these two cases~\footnote{Future neutrinoless
double-beta decay and oscillation experiments will resolve this
ambiguity.}.  As input parameters, we then have the lightest effective
neutrino mass $m_1$ (or $m_3$ for inversely ordered neutrinos), which we
generate in the range $(10^{-4}-0.3)$ eV, the two low scale Majorana
phases $\phi_{1,2}$ and the matrix $H$, which we generate randomly.

We study two different limits of the parameter matrix $H$, of the form
\begin{eqnarray}
H_1=\left(\begin{array}{ccc}
a & 0 & 0 \\
0 & b  & d  \\
0 &  d^\dagger & c
\end{array} \right) \, ,
\label{H1}
\end{eqnarray}
and
\begin{eqnarray}
H_2=\left(\begin{array}{ccc}
a & 0 & d \\
0 & b  & 0  \\
d^\dagger &  0 & c
\end{array} \right) \, ,
\label{H2}
\end{eqnarray}
where $a,b,c$ are real and positive, and $d$ is a complex number.  
We sample these parameters randomly in the range $10^{-2}<a,b,c,|d|<10$,
with distributions that are flat on a logarithmic scale. Also, we 
require the Yukawa coupling-squared to be smaller than $4 \pi$, so that 
$Y_\nu$ remains perturbative up to $M_{G}$. 

\begin{figure}[htbp]
\centerline{\epsfxsize = 0.5\textwidth \epsffile{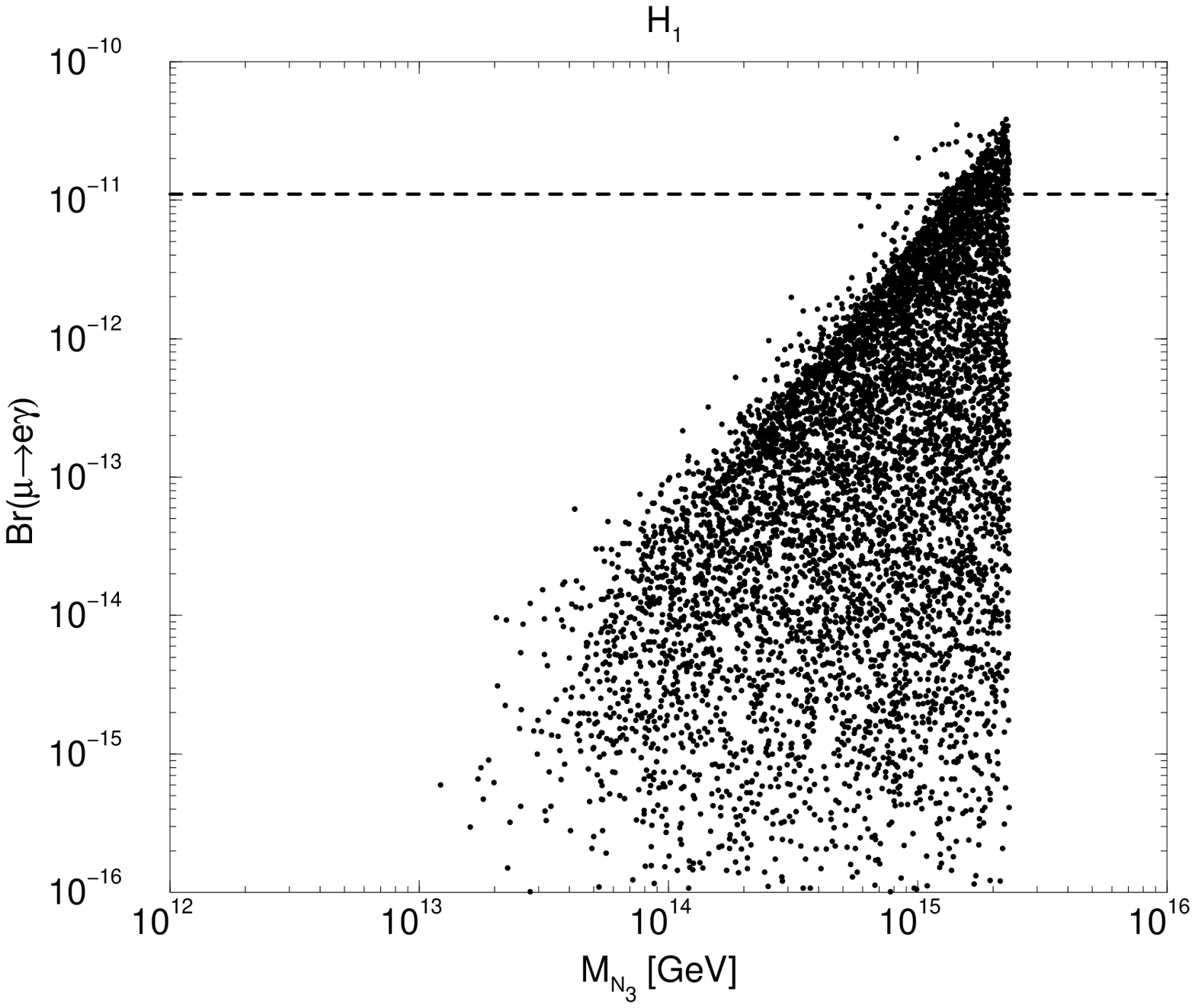} 
\hfill \epsfxsize = 0.5\textwidth \epsffile{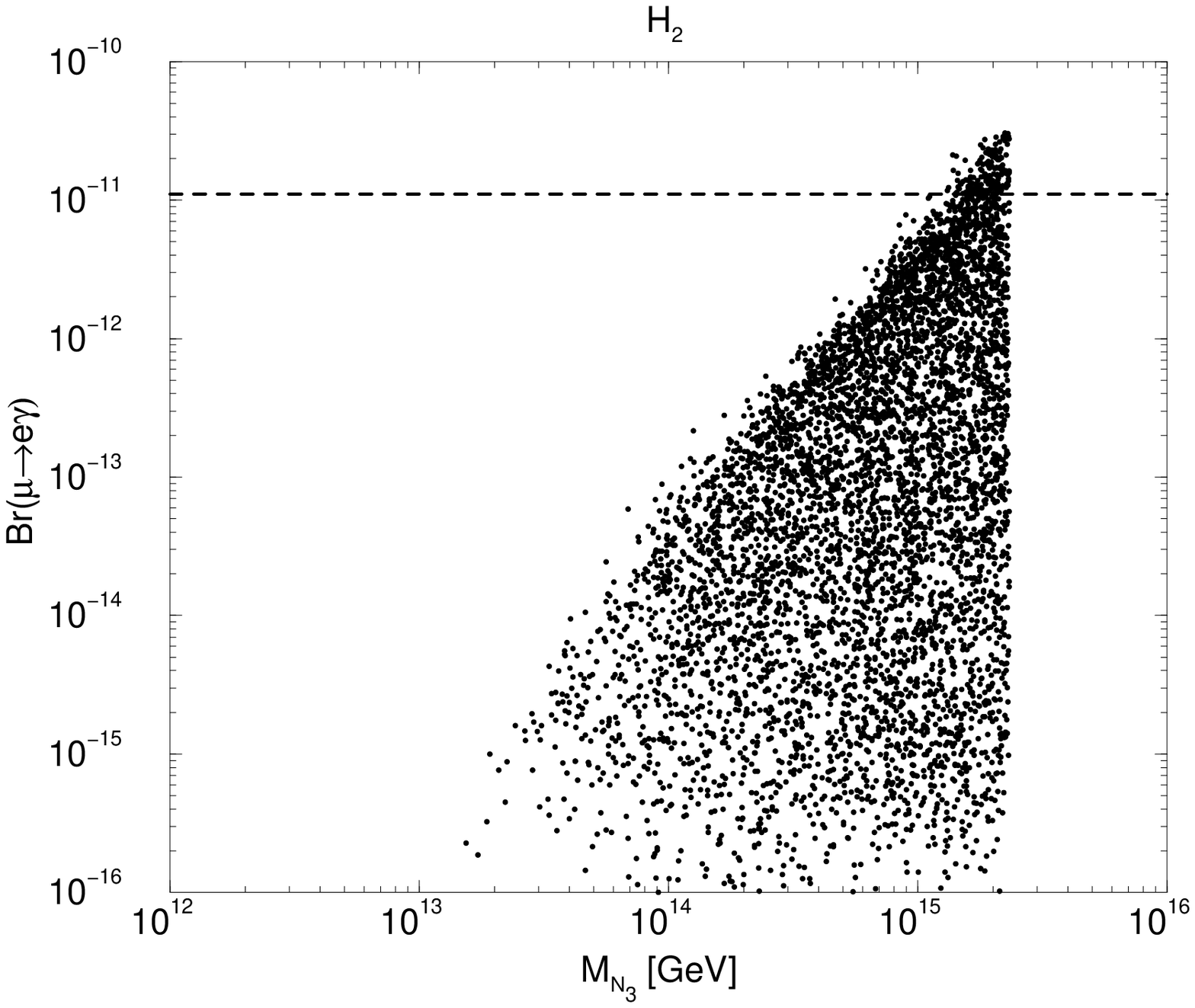} }
\caption{\it
Scatter plot of Br$(\mu\to e\gamma)$ against the heaviest singlet
neutrino mass $M_{N_3}$ for the ansatz (a) $H_1$ and (b) $H_2$. 
We take $m_{1/2}=300$ GeV,
$m_{0}=100$ GeV, $A_{0}=-300$ GeV, $\tan\beta=10$ and $sign(\mu)=+1.$
Other input parameters are specified in the text.
\vspace*{0.5cm}}
\label{fig1}
\end{figure}

In the above ansatz, we take $H_{12}=0$ and $H_{13}H_{32}=0$ because these
conditions suppress Br$(\mu\to e\gamma)$, as seen in (\ref{mueg_const}).  
We show Br$(\mu\to e\gamma)$ as a function of the heaviest right-handed
neutrino mass $M_{N_3}$ for the two structures $H_1$ and
$H_2$ in Fig.~\ref{fig1}. As a reference, here we take $m_{1/2}=300$ GeV,
$m_{0}=100$ GeV, $A_{0}=-300$ GeV, $\tan\beta=10$ and $sign(\mu)=+1.$

With the chosen forms $H_1$ and $H_2$, Br$(\mu\to e\gamma)$ is suppressed
in a broad range of parameters.  In these figures $\mu\rightarrow e
\gamma$ is induced entirely by ${\cal O}(\log M_G/{M}_{N_3}\log{
M}_{N_j}/{M}_{N_i})$ $(i\ne j)$ corrections to the slepton mass matrix and
the slepton trilinear coupling (\ref{thesm},\ref{athre}). Thus, if all
non-zero components in $H_1$ or $H_2$ are of order unity, the correction
to $(m_{\tilde{L}}^2)_{12}$ is not necessary negligible, and one may find
values of the branching ratio above the present experimental bound.  We
also see in Fig.~\ref{fig1} that improving the present sensitivity to
Br$(\mu\to e\gamma)$ by three orders of magnitude, which is currently
being undertaken at PSI~\cite{Barkov}, would be interesting for a large
fraction of the models studied. The MECO and PRISM searches for $\mu^- - 
e^-$
conversion on nuclei~\cite{MECO} would also be interesting in this
respect, and the sensitivities to both processes could be improved at the
front end of a neutrino factory~\cite{nufact}.

The ansatz $H_1$ minimizes $\tau \to e \gamma$ while $\tau\to \mu
\gamma$ can be large, and the opposite is the case for $H_2$, since these
processes are sensitive to  $H_{13}$ and $H_{23}$, respectively. 
Similarly, the EDM of the muon can
be maximized in $H_1$ while the electron EDM can be large in $H_2$. We
exhibit these results in the following subsections.

\subsection{LFV $\tau$ Decay}

\begin{figure}[htbp]
\centerline{\epsfxsize = 0.5\textwidth \epsffile{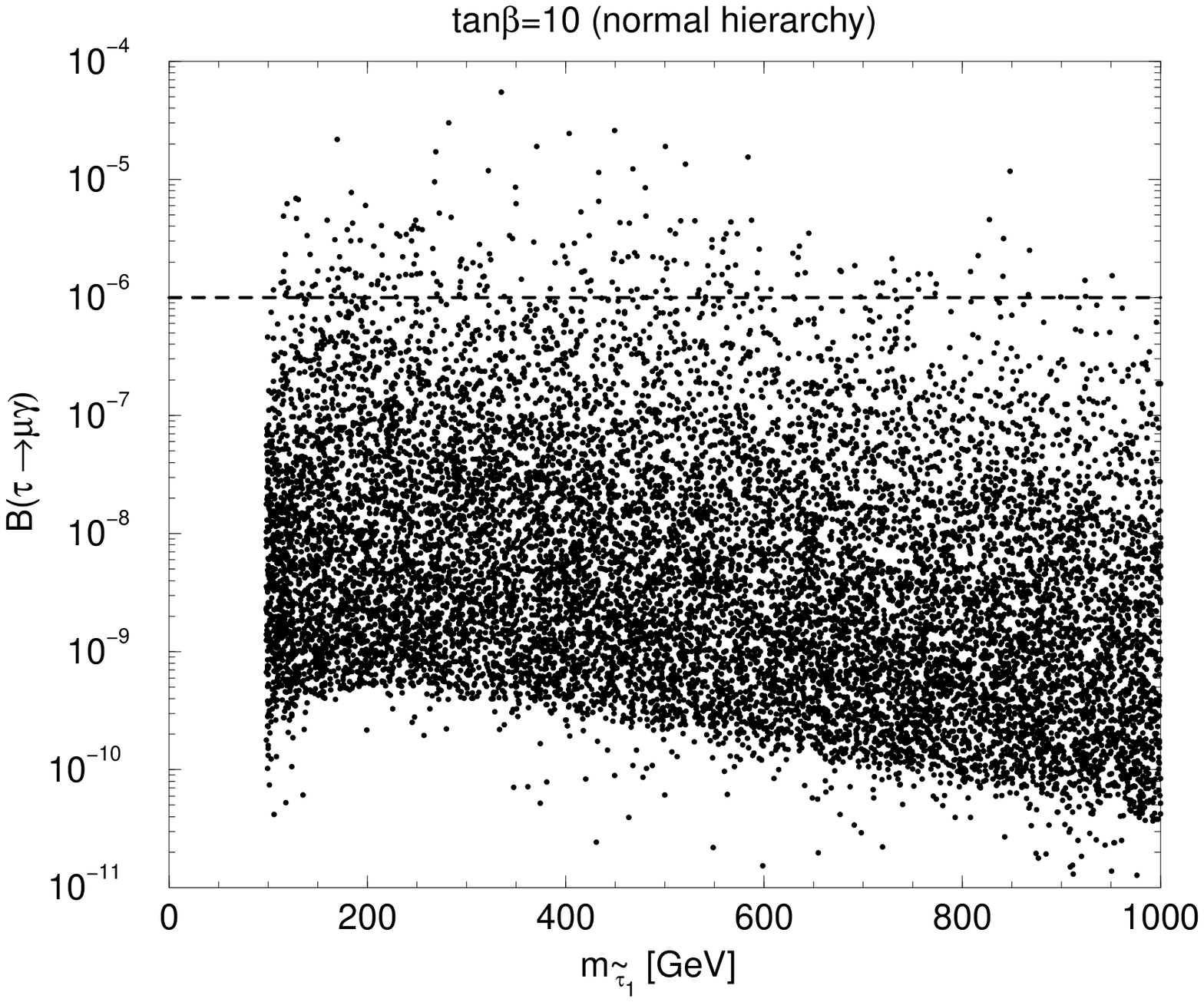} 
\hfill \epsfxsize = 0.5\textwidth \epsffile{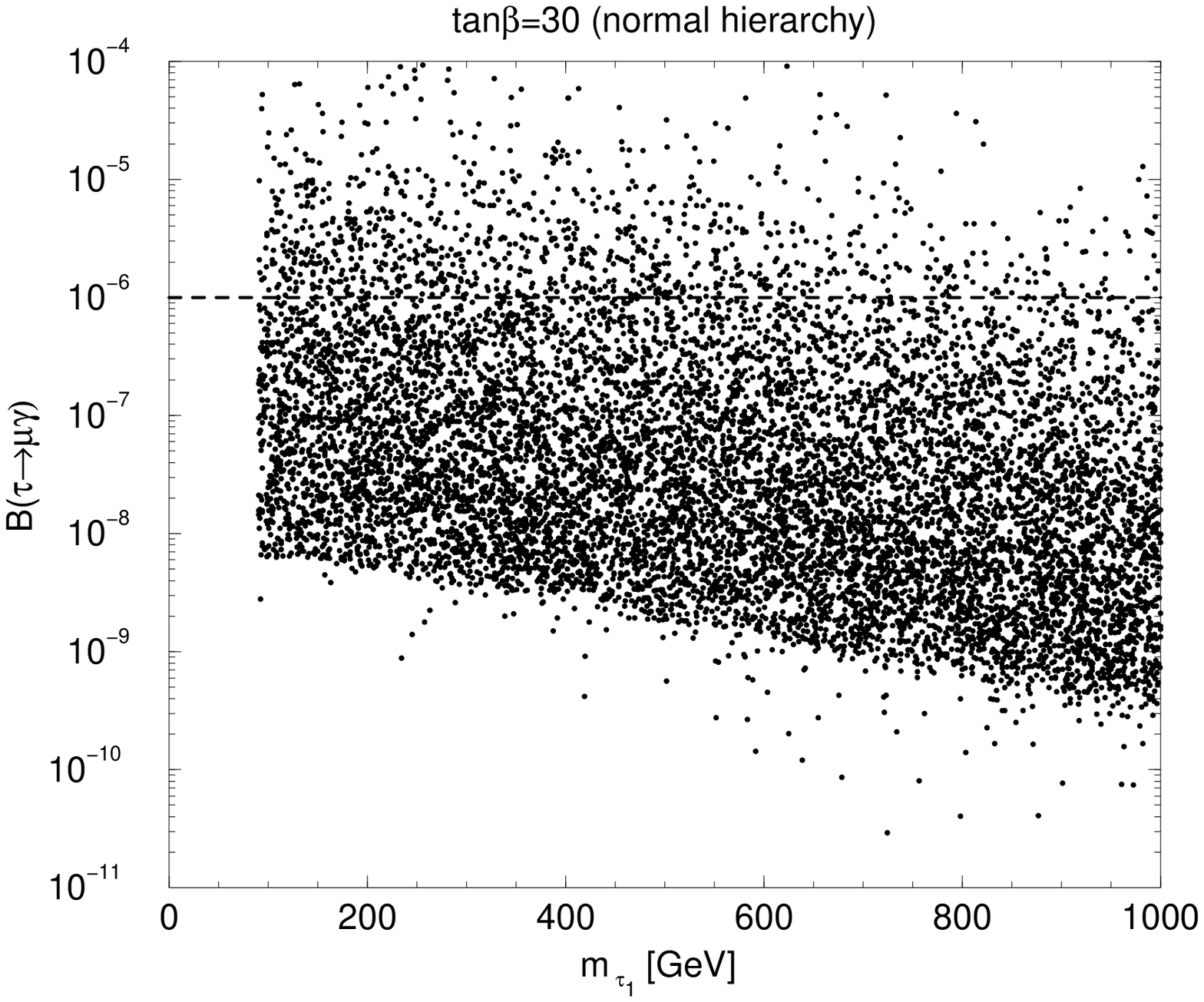} }
\centerline{\epsfxsize = 0.5\textwidth \epsffile{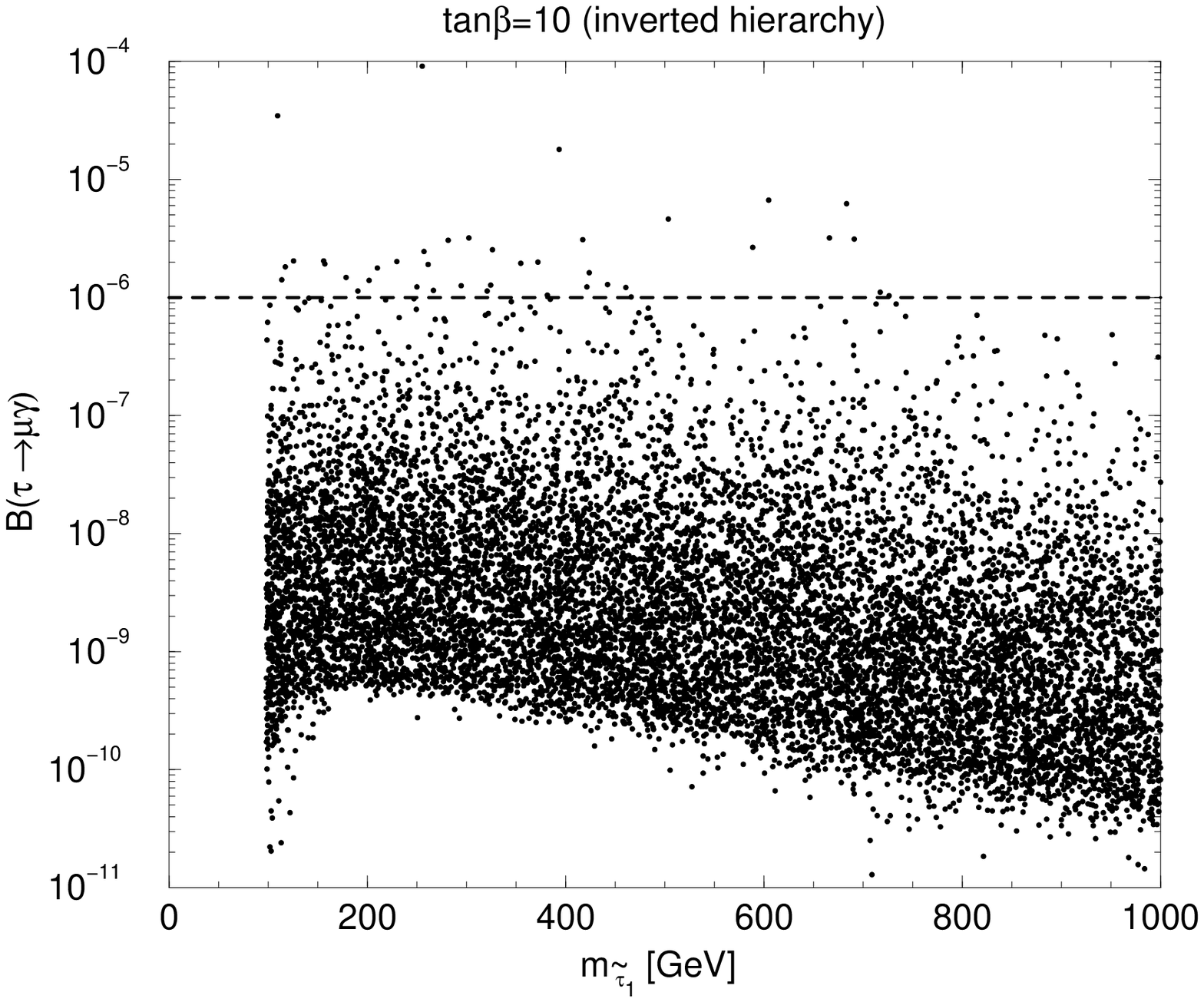} 
\hfill \epsfxsize = 0.5\textwidth \epsffile{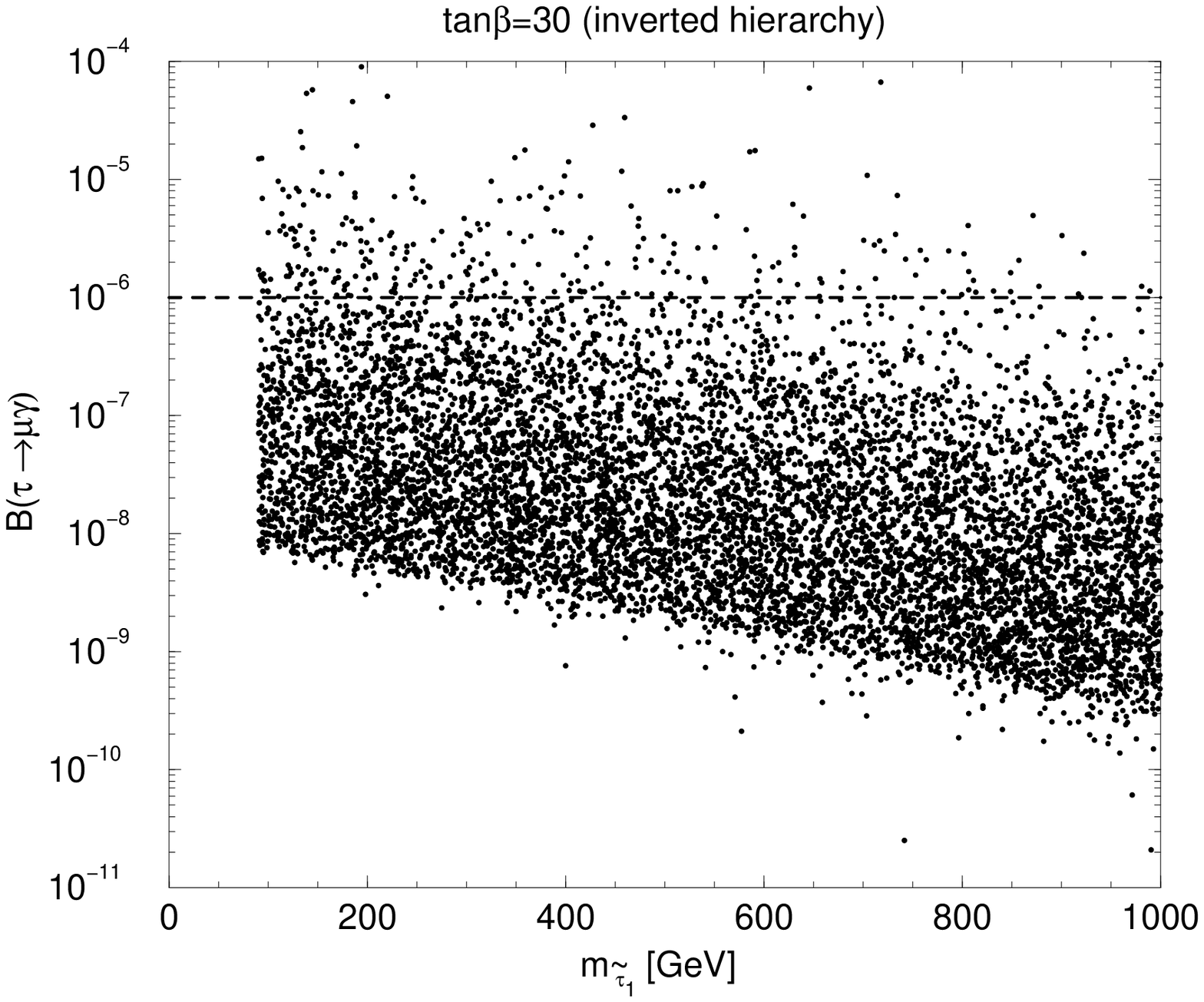} }
\caption{\it 
Scatter plot of Br$(\tau\to \mu \gamma)$ against the lightest stau mass for 
the
ansatz $H_1$. We take the SU(2) gaugino mass to be 200~GeV, $A_0=0$,
$\mu>0$, and $\tan\beta=10$ and $30$. We consider both the normal
and inverted hierarchies for the light neutrino mass spectrum.
\vspace*{0.5cm}}
\label{fig2}
\end{figure}

In Fig.~\ref{fig2} we present Br($\tau\rightarrow \mu \gamma$) for the
ansatz $H_1$, assuming either the normal or the inverted hierarchy for
the light neutrino mass spectrum. The horizontal axis is the lightest
stau mass $m_{\tilde{\tau}_1}$, and the other supersymmetry-breaking
parameters are determined by choosing the SU(2) gaugino mass to be
200~GeV, $A_0=0$, $\mu>0$, and $\tan\beta=10$ and $30$. The parameters
in ${\cal{M}_\nu}$ and $H$ are the same as in Fig.~\ref{fig1}.  The
branching ratio scales as $\tan^2\beta$. We see from these figures
that the branching ratio is similar for the normal and inverted
hierarchies of light-neutrino masses.  In our parametrization the
branching ratio is determined mainly by $H$ and the sparticle mass
spectrum. The dependence on the details of ${\cal M}_\nu$ appears
through the heavy singlet neutrino masses, which influence the
branching ratios only logarithmically.  We find that
Br($\tau\rightarrow \mu \gamma$) can reach even above the present
experimental bound, attaining $10^{-4}$ ($10^{-5}$) for
$\tan\beta=30(10)$, these limits are arising from the perturbative bound on
the neutrino Yukawa coupling. 

\begin{figure}[htbp]
\centerline{\epsfxsize = 0.5\textwidth \epsffile{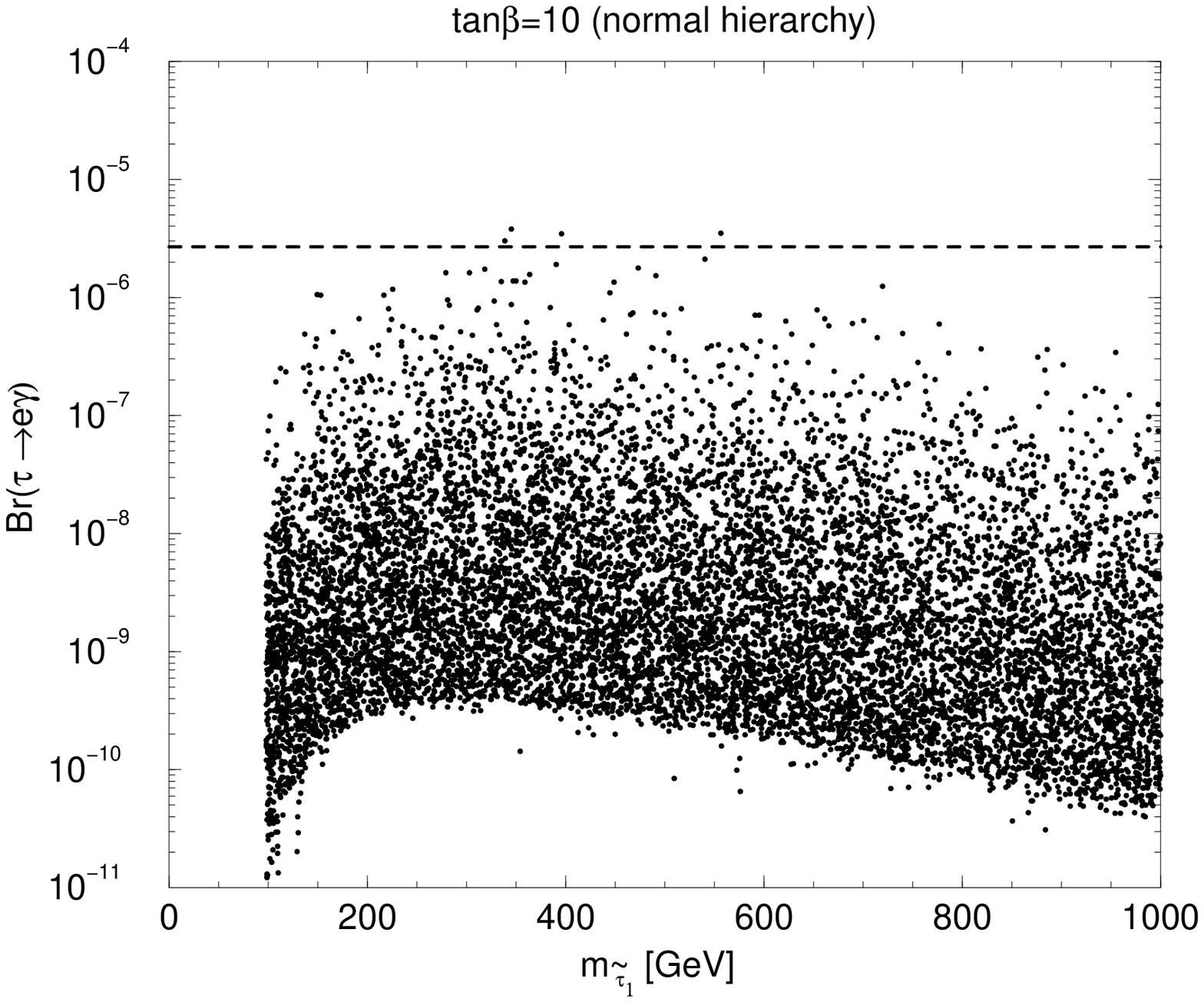} 
\hfill \epsfxsize = 0.5\textwidth \epsffile{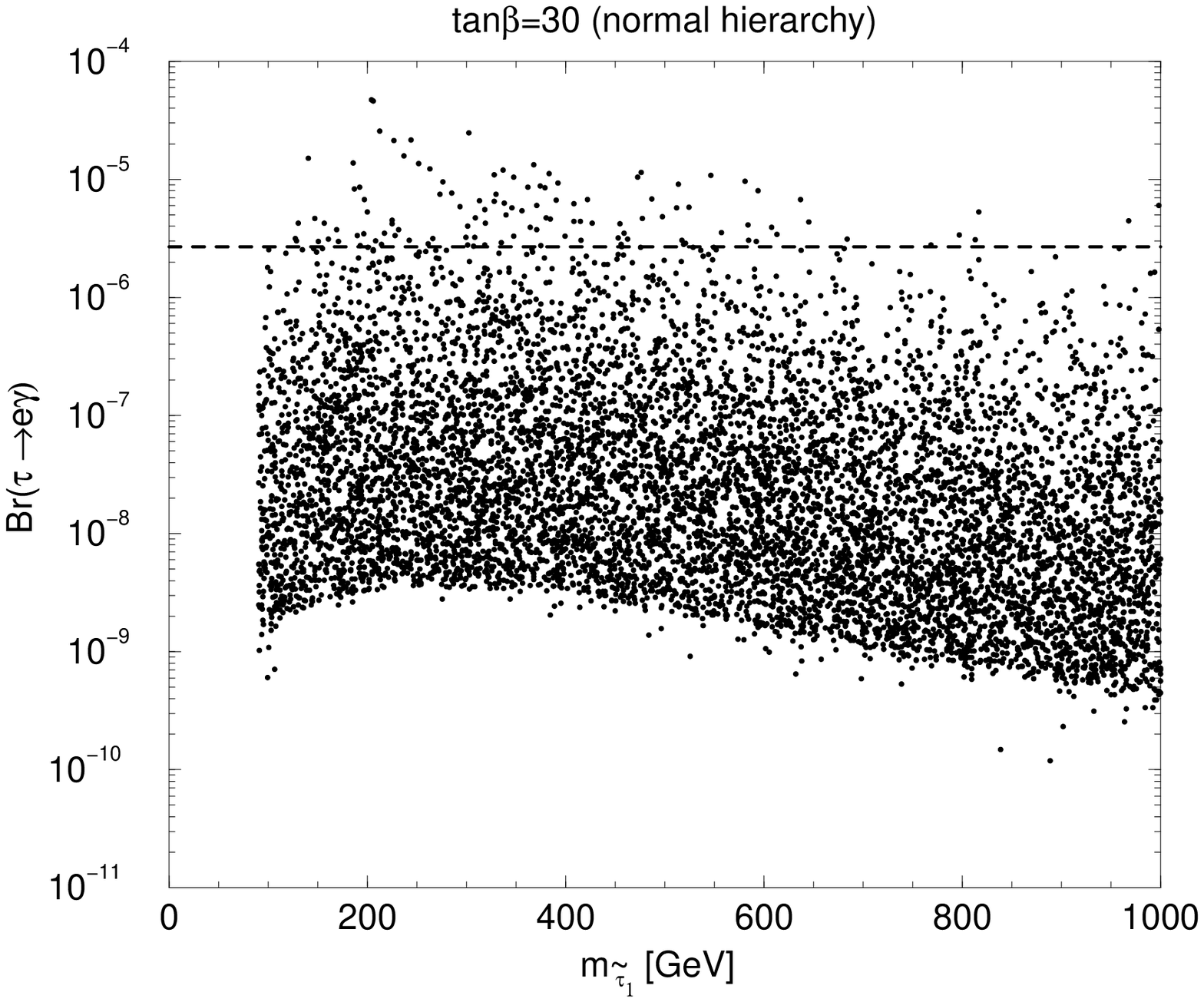} }
\centerline{\epsfxsize = 0.5\textwidth \epsffile{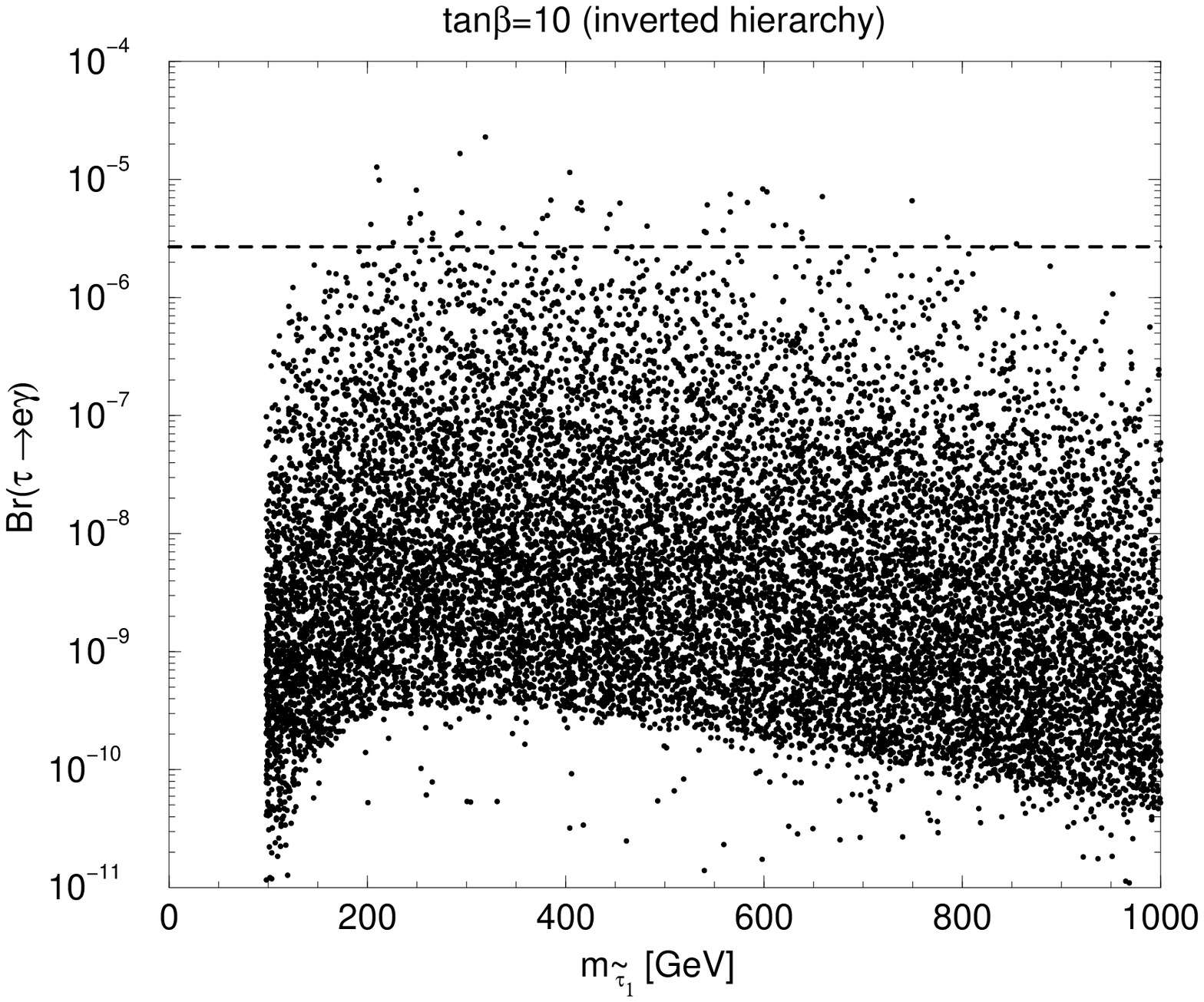} 
\hfill \epsfxsize = 0.5\textwidth \epsffile{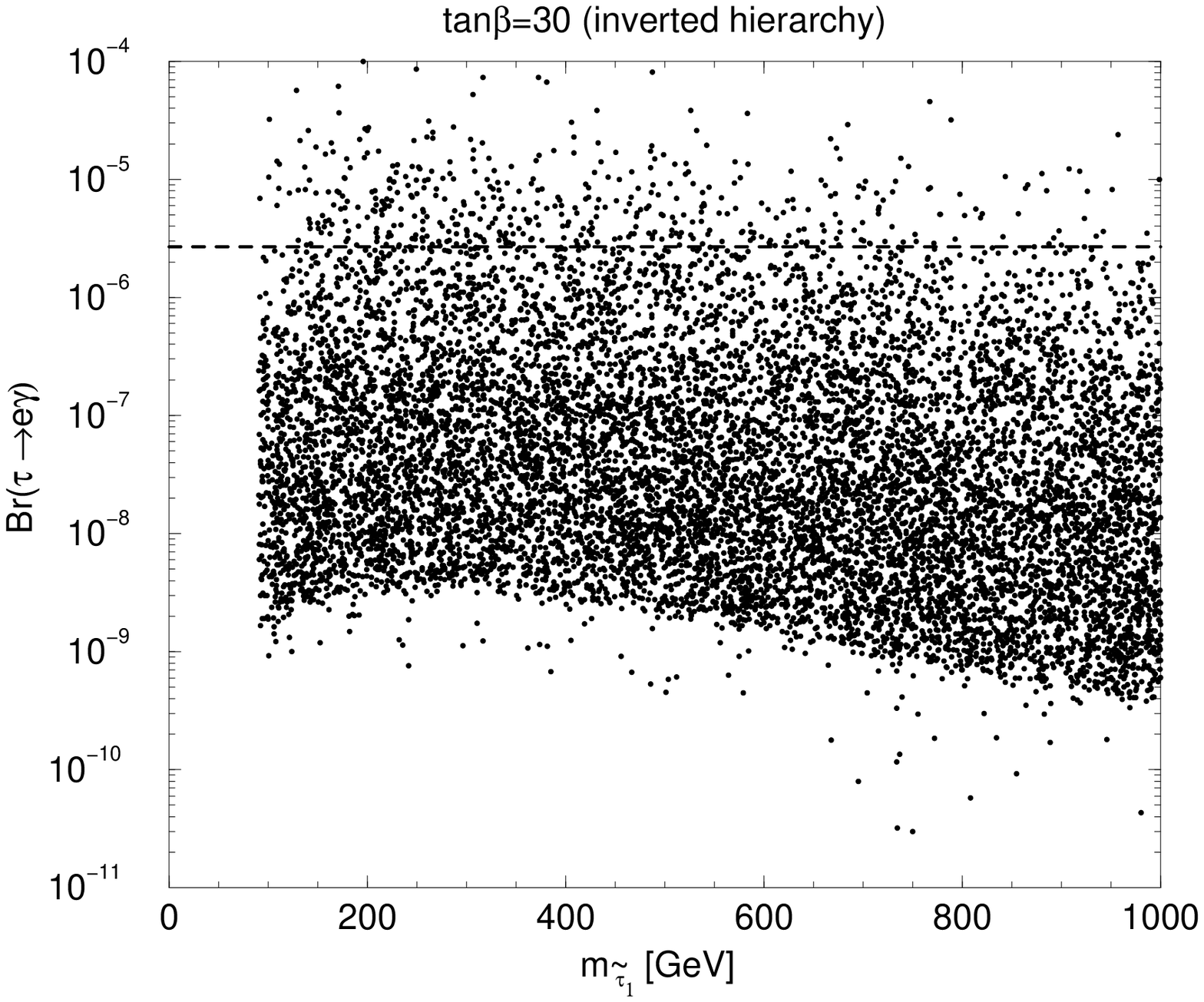} }
\caption{\it 
Scatter plot of Br$(\tau\to e \gamma)$ against the lightest stau 
mass for the ansatz $H_2$. 
The input parameters for the supersymmetry-breaking parameters are
the same as in Fig.~\ref{fig2}.
\vspace*{0.5cm}}
\label{fig3}
\end{figure}

Next we show Br($\tau\rightarrow e \gamma$) for the ansatz $H_2$ in
Fig.~\ref{fig3}. The input parameters are the same as in Fig.~\ref{fig2},
and the behaviour is similar to that of Br($\tau\rightarrow \mu \gamma$) in
Fig.~\ref{fig2}. This process may also reach to $10^{-4}$ ($10^{-5}$) for
$\tan\beta=30(10)$.

We see in Fig.~\ref{fig2},~\ref{fig3} that improving the 
present sensitivity to Br$(\tau\to \mu/e \gamma)$ by two orders of
magnitude, which seems feasible with the present generation of B 
factories~\cite{ohshima}, would be very interesting for many of the models 
studied.
Whilst the B factories have good sensitivities to $\tau\rightarrow \mu 
\gamma$ and $\tau\rightarrow e \gamma$, hadron colliders may have better
reach for $\tau\rightarrow 3\ell$ decays. We
present here a simple formula for the LFV $\tau$ decays to three charged
leptons.  When $(m^2_{\tilde{L}})_{23}$ is non-vanishing, non-zero
$\tau\rightarrow \mu\gamma$ and $\tau\rightarrow \mu \ell^+\ell^-$ are 
both predicted. The
photonic penguin diagram tends to dominate over other contributions in 
the trilepton final state due to the phase-space integral. When
$\tan\beta$ is large, the dominance is even stronger. When the
photonic penguin diagram is dominant in $\tau\rightarrow \mu 
\ell^+\ell^-$,
\begin{eqnarray}
\frac{{\rm Br}(\tau\rightarrow 3\mu)}
    {{\rm Br}(\tau\rightarrow \mu \gamma)}
&=&
\frac{\alpha}{8\pi}\frac{8}{3} 
\left(\log\frac{m_\tau^2}{m_\mu^2}-\frac{11}4\right)
\simeq \frac{1}{440}\,,
\\
\frac{{\rm Br}(\tau\rightarrow \mu 2 e )}
    {{\rm Br}(\tau\rightarrow \mu \gamma)}
&=&
\frac{\alpha}{8\pi}\frac{8}{3} 
\left(\log\frac{m_\tau^2}{m_e^2}-\frac{8}{3}\right)
\simeq
\frac1{94}\,.
\end{eqnarray}
The branching ratio for $\mu e^+ e^-$ is larger than 
that to $\mu\mu^+\mu^-$, because the phase space is larger. 
Similarly, we get the following relations for the $\tau$--$e$ transition:
\begin{eqnarray}
\frac{{\rm Br}(\tau\rightarrow 3e)}
    {{\rm Br}(\tau\rightarrow e \gamma)}
&\simeq&
\frac{1}{95}\,,
\\
\frac{{\rm Br}(\tau\rightarrow e 2\mu )}
    {{\rm Br}(\tau\rightarrow e \gamma)}
&\simeq&
\frac1{430}\,.
\end{eqnarray}
Thus, from the experimental bound on ${\rm Br}(\tau\rightarrow\mu/e
\gamma)$, we find that ${\rm Br}(\tau\rightarrow \mu 2 e)$ and 
${\rm Br}(\tau\rightarrow 3 e)$ can reach $10^{-8}$, and ${\rm
Br}(\tau\rightarrow e 2\mu)$ and ${\rm Br}(\tau\rightarrow 3\mu)$ can
reach $10^{-9}$. 

\subsection{EDMs of Charged Leptons}

The EDMs of charged leptons are induced essentially through the threshold
correction to $A_e$ at the heavy singlet-neutrino scale, and the
dependence on $H$ and ${\cal M}_\nu$ is complicated. From (\ref{athre}) it
is found that the imaginary parts of diagonal terms in $A_e$ can reach
${\cal O}(0.1)$\% if the $(X_k)_{ij}$ are no larger than $4\pi$. 
In this case, the
muon and electron EDMs can roughly reach the level of $10^{-25}e~{\rm cm}$
and $10^{-27} e~{\rm cm},$ respectively.  However, the non-negligible
contribution of the heavy singlet neutrino threshold correction may lead
to large Br($\mu\rightarrow e \gamma$) even if $H_{12}$ and $H_{13}H_{32}$
are suppressed.  Therefore careful numerical study is required for the
predictions of EDMs in the supersymmetric seesaw model.

\begin{figure}[htbp]
\centerline{\epsfxsize = 0.5\textwidth \epsffile{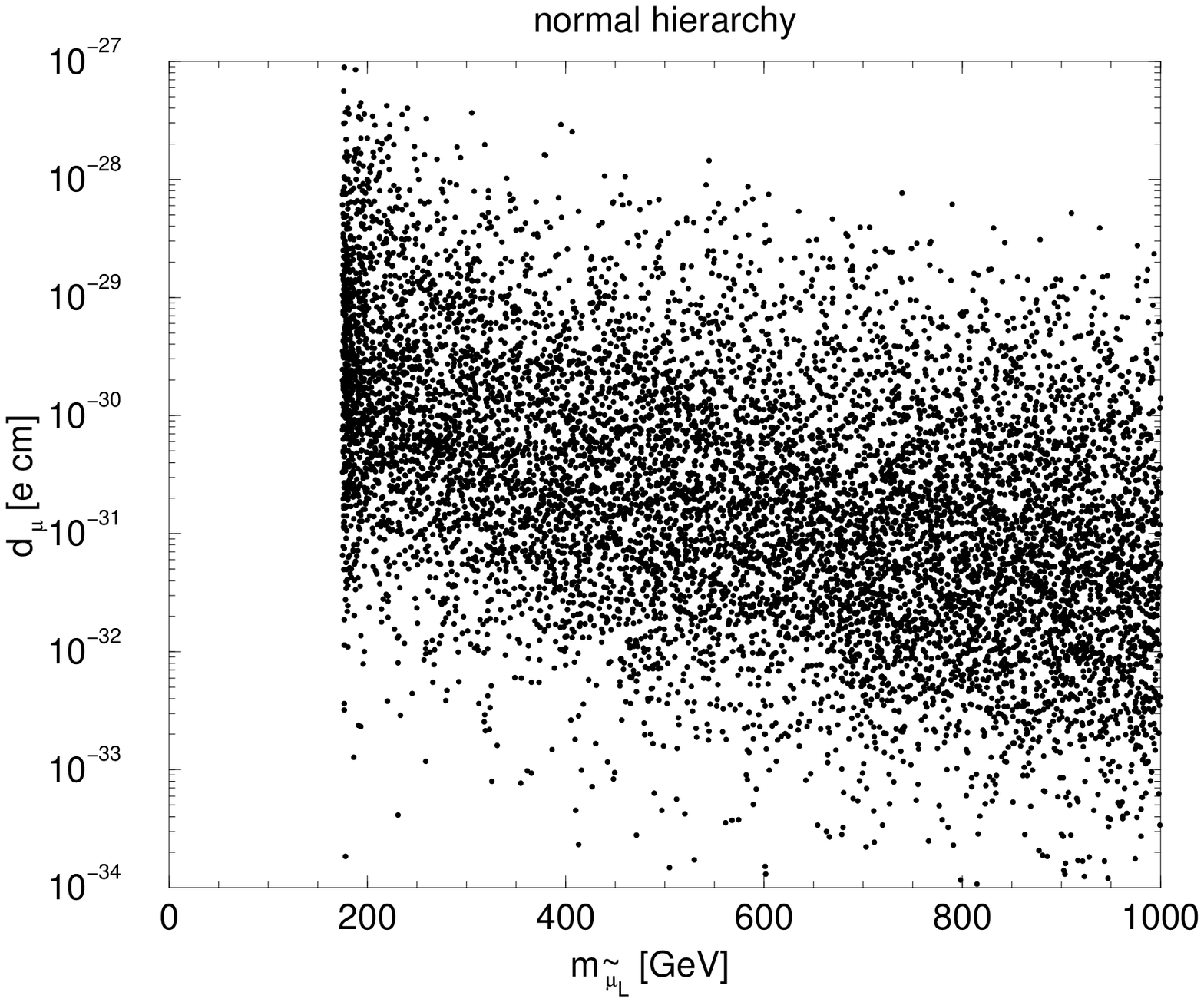} 
\hfill \epsfxsize = 0.5\textwidth \epsffile{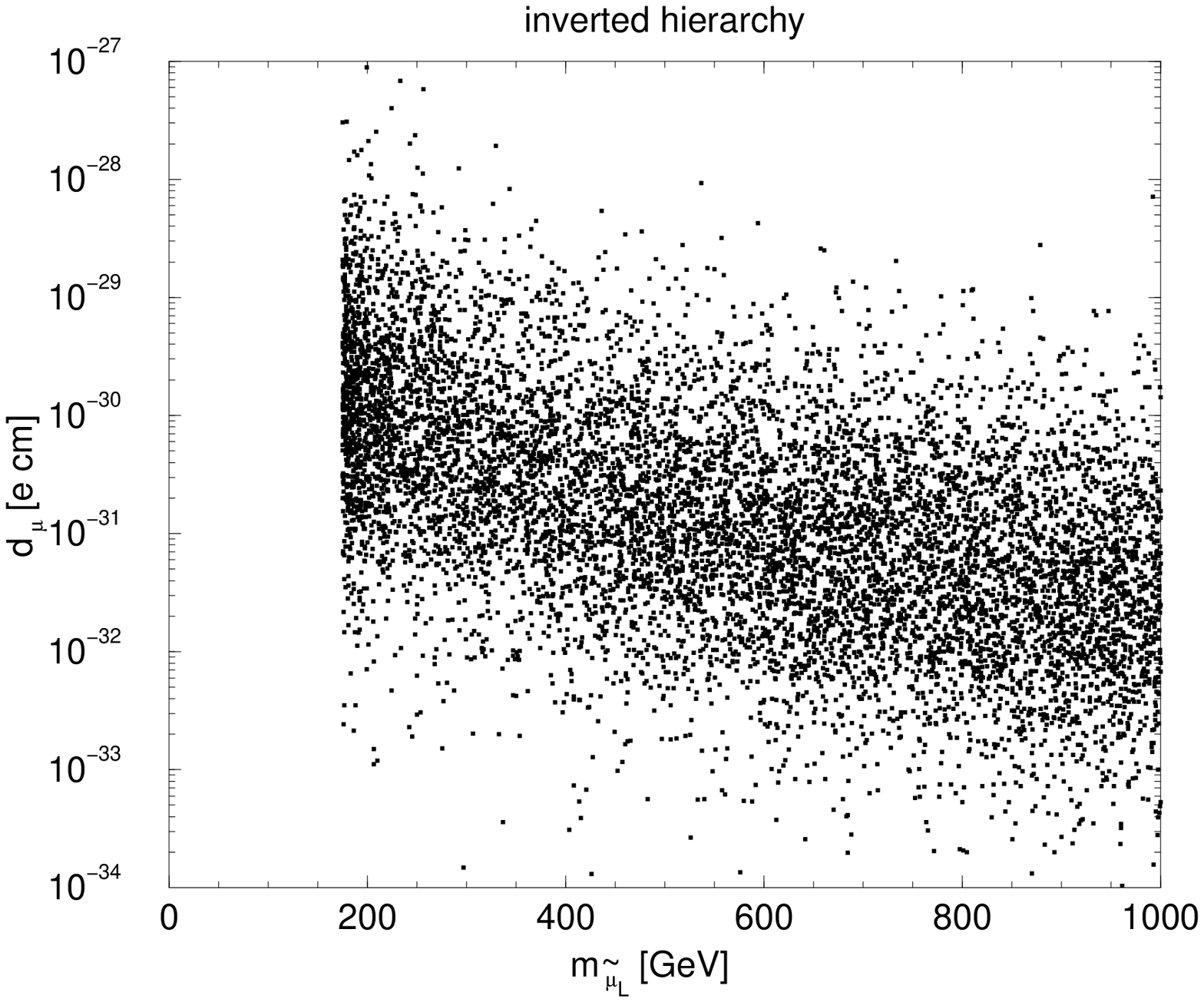}}
\caption{\it 
Scatter plot of the muon EDM against the left-smuon mass for the 
ansatz $H_1$, taking the SU(2) gaugino mass to be 200~GeV, $A_0=-3m_0$, 
$\mu>0$, and
$\tan\beta=10$. We assume the normal hierarchy for the light neutrino mass
spectrum in (a) and the inverted hierarchy in (b).
\vspace*{0.5cm}}
\label{fig4}
\end{figure}

We show first in Fig.~\ref{fig4} predictions for the EDMs of the muon and
electron from a random sampling of the parameter space, as a scatter
plot of the muon EDM against the left-smuon mass for the ansatz $H_1$.
As before, we take the SU(2) gaugino mass to be 200~GeV, $A_0=-3m_0$,
$\mu>0$, and $\tan\beta=10$. The parameters in ${\cal{M}_\nu}$ and $H$ are 
the same as in Fig.~\ref{fig1}. We assume the normal hierarchy for the light
neutrino mass spectrum, and impose the constraints from Br($\mu\rightarrow
e \gamma$) and Br($\tau\rightarrow\mu(e) \gamma$). We find that the muon 
EDM can reach $10^{-(27-28)}e$~cm in this
sampling, which is reasonable, since $(m_{\tilde{L}}^2)_{12}$ should
suppressed by $10^{-(3-4)}$ because of the experimental bound on
Br($\mu\rightarrow e \gamma$). The EDM increases proportionally to $A_0$,
as expected, but is insensitive to $\tan\beta$, if it is not
extremely large.

\begin{figure}[htbp]
\centerline{\epsfxsize = 0.5\textwidth \epsffile{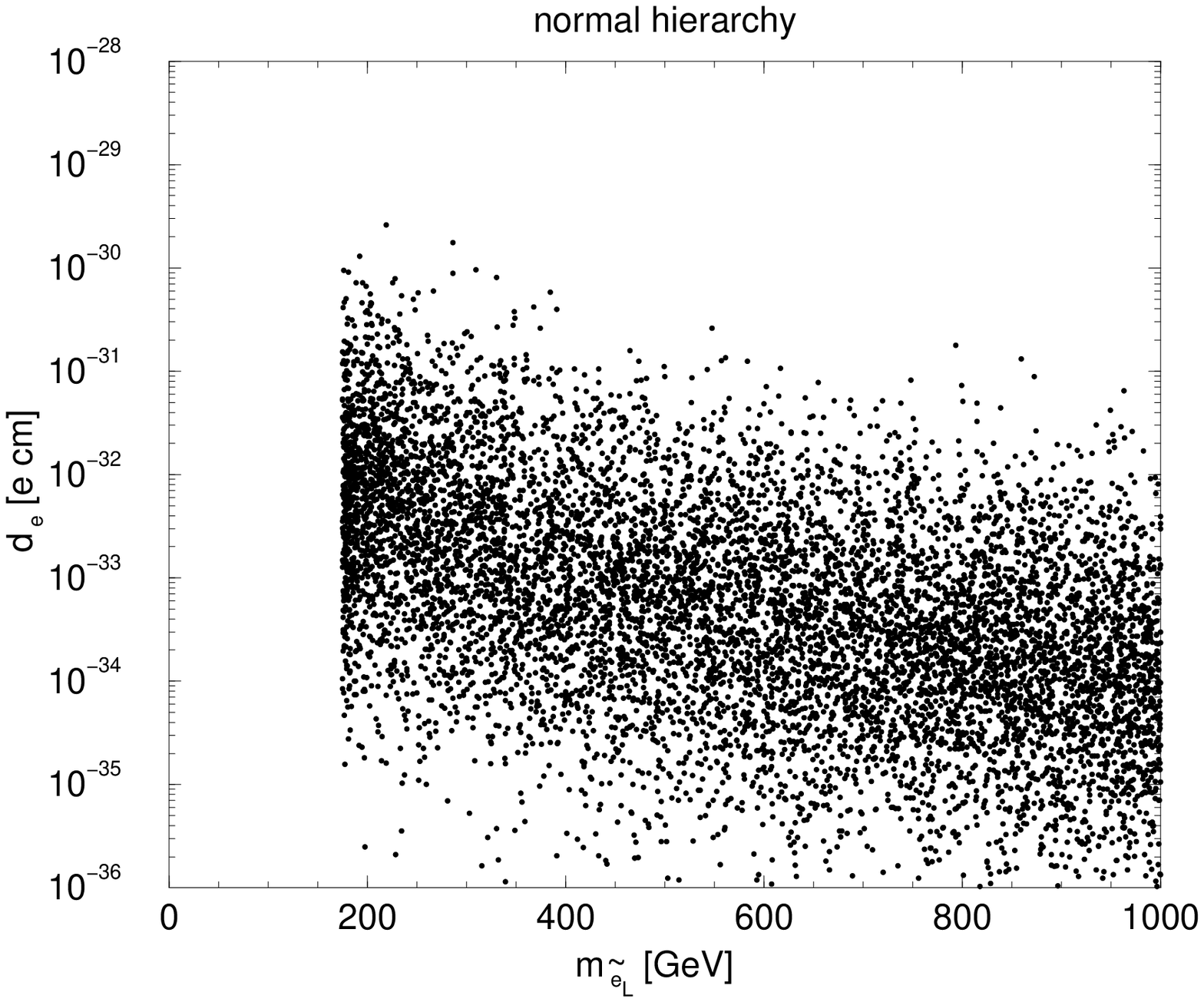} 
\hfill \epsfxsize = 0.5\textwidth \epsffile{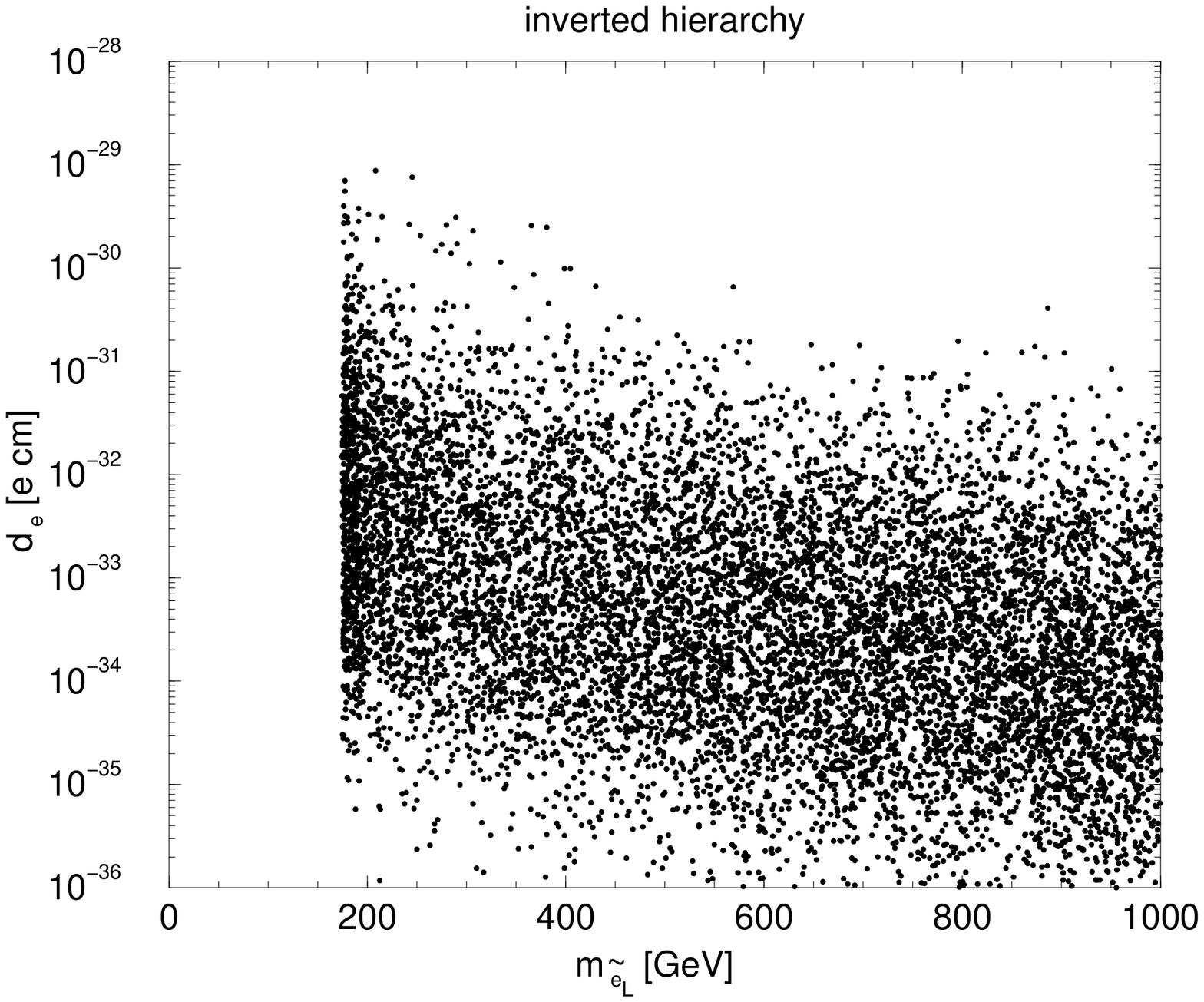}}
\caption{\it 
Scatter plot of the electron EDM against the left-selectron mass 
for the ansatz $H_2$. Other input parameters are the same as in
Fig.~\ref{fig4}.
\vspace*{0.5cm}}
\label{fig5}
\end{figure}

Similarly, in Fig.~\ref{fig5} we show the electron EDM plotted against the
left-selectron mass for the ansatz $H_2$. We find that the 
electron EDM can reach $10^{-(29-30)}e$~cm in this sampling.
Again, the main limiting factor is the experimental
bound on Br($\mu\rightarrow e \gamma$).

We recall that a proposal has been made to BNL that aims at a sensitivity
of $10^{-24}e$~cm for the muon EDM~\cite{BNL}, and the front end of a
neutrino factory may be able to reach a sensitivity of $5 \times
10^{-26}e$~cm~\cite{nufact}. On the other hand, although the present upper
limit on the electron EDM is $1.6 \times 10^{-27}e$~cm~\cite{eEDM}, a
technique has been proposed that may be sensitive to
$10^{-32}e$~cm~\cite{Lam}. This would be sensitive to many of the models
studied.

\section{Conclusions}

We have presented a new parametrization of the minimal seesaw mechanism,
which enables the heavy neutrino Dirac Yukawa couplings and masses to be
fixed in terms of the light neutrino parameters and a Hermitian parameter
matrix $H.$ In the minimal supersymmetric version of the seesaw model, the
matrix $H$ can be related directly to low-energy physical observables. As
a result, our parametrization is particularly suitable for comprehensive
studies of the charged LFV processes and EDMs in supersymmetric models.

As applications, we have studied the LFV $\tau$ decays and the EDMs of
the muon and electron in the minimal supersymmetric seesaw model. It
is found that Br($\tau\rightarrow \mu(e)\gamma$) could exceed the
present experimental bounds, even when Br($\mu\rightarrow e\gamma$) is
suppressed much below the current limit. This implies that B factories
have a possibility of discovering LFV $\tau$ decays, since they may
reach sensitivities Br($\tau\rightarrow \mu(e)\gamma$)$\sim
10^{-8}$. The LHC may have a similar sensitivity, and a super B
factory may reach the level $10^{-9}$ for the same
processes~\cite{ohshima}.  The LHC may also have a good sensitivity to
Br($\tau\rightarrow \mu(e)\ell^+\ell^-$). We show that
Br($\tau\rightarrow \mu ee$) and Br($\tau\rightarrow 3e $) are about
five times larger than Br($\tau\rightarrow 3\mu$) and
Br($\tau\rightarrow e 2 \mu $), due to the larger phase space, and can
reach $\sim 10^{-8}$ and $\sim 10^{-9}$, respectively, from the
experimental bounds on Br($\tau\rightarrow \mu(e)\gamma$).  Finally,
in our random samples the EDMs of muon and electron can attain
$10^{-(27-28)}e$~cm and $10^{-(29-30)}e$~cm, respectively, while their
perturbative bounds in this model are $\sim 10^{-25}e$~cm and $\sim
10^{-27}e$~cm. The electron EDM, in particular, may be accessible to 
experiment~\cite{Lam}.

We have restricted our discussion to the supersymmetric seesaw model in
this paper. However, our framework is also suitable for studying
supersymmetric GUT models with heavy singlet neutrinos. In these models,
$K$ and $B$ physics are also interesting, because the right-handed squarks
have flavour-violating masses, as a result of quark-lepton
unification~\cite{susyguthadron}.  Since the right-handed sleptons have
LFV masses, the relation between the LFV and our parametrization may be
quite complicated~\cite{sglfv}.  However, if the prediction for the LFV
masses for right-handed sleptons is used, our parametrization is
applicable.

\vskip 0.5in
\vbox{
\noindent{ {\bf Acknowledgments} } \\
\noindent  
We thank C. Pe\~na-Garay and A. Strumia for discussions.
This work is partially supported by EU TMR
contract No.  HPMF-CT-2000-00460, ESF grant No. 5135, the
Grant-in-Aid for Scientific Research from the Ministry of Education,
Science, Sports and Culture of Japan (No. 13135207 and No. 14046225),
and the JSPS fellowship.
}

\end{document}